\documentclass[11pt,fleqn]{article}

\usepackage{a4}
\usepackage{amstext}  
\usepackage{amsfonts}  
\usepackage{amssymb}  
\usepackage{color}
\usepackage{epsfig}
\usepackage{macros}

\newcommand{\ltapprox}{\raisebox{-0.5ex}{$\,\stackrel{<}{\scriptstyle\sim}\,$}}

\begin{document}


\begin{flushright}
SFB/CPP-08-82 \\ DESY 08-135 \\ LTH 809 \\ HU-EP-08/44
\end{flushright}

\begin{center}

{\huge \bf The static-light meson spectrum \\ from twisted mass lattice QCD}

\vspace{0.5cm}

\textbf{Karl Jansen} \\
DESY, Platanenallee 6, D-15738 Zeuthen, Germany \\
\texttt{karl.jansen@desy.de}

\vspace{0.5cm}

\textbf{Chris Michael, Andrea Shindler} \\
Theoretical Physics Division, Department of Mathematical Sciences, University of Liverpool, Liverpool L69 3BX, UK \\
\texttt{c.michael@liverpool.ac.uk} \\
\texttt{andrea.shindler@liverpool.ac.uk}

\vspace{0.5cm}

\textbf{Marc Wagner} \\
Humboldt-Universit\"at zu Berlin, Institut f\"ur Physik, Newtonstra{\ss}e 15, D-12489 Berlin, Germany  \\
\texttt{mcwagner@physik.hu-berlin.de}

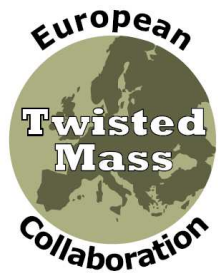
\vspace{0.5cm}

\begin{picture}(0,0)%
\includegraphics{Logo.pstex}%
\end{picture}%
\setlength{\unitlength}{4144sp}%
\begingroup\makeatletter\ifx\SetFigFont\undefined%
\gdef\SetFigFont#1#2#3#4#5{%
  \reset@font\fontsize{#1}{#2pt}%
  \fontfamily{#3}\fontseries{#4}\fontshape{#5}%
  \selectfont}%
\fi\endgroup%
\begin{picture}(1620,1620)(1,-781)
\end{picture}%

\vspace{0.4cm}

October 9, 2008

\end{center}

\vspace{0.1cm}

\begin{tabular*}{16cm}{l@{\extracolsep{\fill}}r} \hline \end{tabular*}

\vspace{-0.4cm}
\begin{center} \textbf{Abstract} \end{center}
\vspace{-0.4cm}

We compute the static-light meson spectrum with $N_f = 2$ flavours of sea quarks using Wilson twisted mass lattice QCD.  We consider five different values for the light quark mass corresponding to $300 \, \textrm{MeV} \ltapprox m_\mathrm{PS} \ltapprox 600 \, \textrm{MeV}$ and we present results for angular momentum $j = 1/2$, $j = 3/2$ and $j = 5/2$ and for parity $\mathcal{P} = +$ and $\mathcal{P} = -$. We extrapolate our results to physical quark masses and make predictions regarding the spectrum of $B$ and $B_s$ mesons.

\begin{tabular*}{16cm}{l@{\extracolsep{\fill}}r} \hline \end{tabular*}

\thispagestyle{empty}


\newpage

\setcounter{page}{1}

\section{\label{SEC476}Introduction}

A systematic way to study $B$ and $B_s$ mesons from first principles is
with lattice QCD.  Since $a m_b > 1$ at currently available lattice
spacings for large volume simulations,  one needs to use for the $b$
quark a formalism such as Heavy Quark Effective Theory (HQET)  or
Non-Relativistic QCD. Here we follow the HQET route,  which enables all
sources of systematic error to be controlled.

In the static limit a heavy-light meson will be the ``hydrogen atom'' of QCD. Since in this limit there are no interactions involving the heavy quark spin, states are doubly degenerate, i.e.\ there is no hyperfine splitting. Therefore, it is common to label static-light mesons by parity $\mathcal{P}$ and total angular momentum of the light degrees of freedom $j$ with $j= |l \pm 1/2|$, where $l$ denotes angular momentum and $\pm 1/2$ the spin of the light quark. An equivalent notation is given by $S \equiv (1/2)^-$, $P_- \equiv (1/2)^+$, $P_+ \equiv (3/2)^+$, $D_- \equiv (3/2)^-$, ... The total angular momentum of the static-light meson is either $J = j + 1/2$ or $J = j - 1/2$, where both states are of the same mass. Note that in contrast to parity, charge conjugation is not a good quantum number, since static-light mesons are made from non-identical quarks.

The static-light meson spectrum has been studied comprehensively by lattice methods in the quenched approximation with a rather coarse lattice spacing \cite{Michael:1998sg}. Lattice studies with $N_f=2$ flavours of dynamical sea quarks have also explored this spectrum \cite{Green:2003zza,Burch:2006mb,Koponen:2007fe,Foley:2007ui,Koponen:2007nr,Burch:2008qx}. Here (cf.\ also \cite{Jansen:2008ht}) we use $N_f=2$ and are able to reach lighter dynamical quark masses, which are closer to the physical $u/d$ quark mass, so enabling a more reliable extrapolation. Note that in this formalism, mass differences in the heavy-light spectrum are $\mathcal{O}(a)$ improved so that the continuum limit is more readily accessible.

 In this paper, we concentrate on the unitary sector, where valence
quarks and sea quarks are of the same mass.  This is appropriate for
static-light mesons with a light quark, which is $u/d$. We also estimate
masses of static-light mesons with light $s$ quarks, albeit with a sea
of two degenerate $s$ instead of $u$ and $d$.
 Within the twisted mass formalism, it is feasible to use $N_f = 2+1+1$
flavours of dynamical sea quarks, which will give a more appropriate
focus on the static-strange meson spectrum with light sea quarks.

In HQET the leading order is just the static limit. The next correction
will be of order $1/m_Q$, where $m_Q$ is the mass of the heavy quark.
This correction is expected be relatively small for $b$ quarks, but
larger for $c$ quarks. Lattice methods to evaluate these $1/m_Q$
contributions to the $B$ meson hyperfine splittings have been
established and tested in quenched studies
\cite{Bochicchio:1991cy,Guazzini:2007bu}. We intend to explore these
contributions using lattice techniques subsequently. An alternative way
to predict the spectrum for $B$ and $B_s$ mesons is to interpolate
between $D$ and $D_s$ states, where the experimental spectrum is rather
well known, and the static limit obtained by lattice QCD assuming a
dependence as $1/m_Q$. Thus the splittings among $B$ and $B_s$ mesons
should be approximately $m_c / m_b \approx 1/3$ of those among the
corresponding $D$ and $D_s$ mesons.

For excited $D_s$ mesons, experiment has shown that some of the states
have very narrow decay widths \cite{PDG}. This comes about, since the
hadronic transitions to $D K$ and $D_s M$ (where $M$ is a flavour
singlet mesonic system, e.g.\ $\eta'$, $\pi \pi$ or $f_0$) are not
allowed energetically. The isospin violating decay to $D_s \pi$ together
with electromagnetic decay to $D_s \gamma$  are then responsible for the
narrow width observed. A similar situation may exist for $B_s$ decays
and we investigate this here using our lattice mass determinations of
the excited states. This will enable us to predict whether narrow
excited $B_s$ mesons should be found. 

As well as exploring this issue of great interest to experiment, we
determine the excited state spectrum  of static-light mesons as fully as
possible. This will help the construction of phenomenological models and
will shed light on questions such as, whether there is an inversion of
the level ordering with $l_+$ lighter than $l_-$ at larger $l$ or for
radial excitations as has been predicted
\cite{Schnitzer:1978gq,Schnitzer:1989xr,Ebert:1997nk,Isgur:1998kr}.

Since we measure the spectrum for a range of values of the bare quark
mass parameter  $\mu_\mathrm{q}$ for the light quark, we could also
compare with chiral effective Lagrangians   appropriate to HQET. This
comparison would be most appropriate applied to heavy-light decay
constants in the continuum limit,  so we will defer that discussion to a
subsequent publication.   

This paper is organised as follows. In section~\ref{SEC466} we review
some basic properties of twisted mass lattice QCD. Moreover, we discuss
particularities arising in static-light computations as well as
automatic $\mathcal{O}(a)$ improvement. In section~\ref{SEC399} we
present technical details regarding static-light meson creation
operators and the corresponding correlation matrices we are using. We
also explain how we extract the static-light spectrum from these
correlation matrices and present numerical results for a range of light
quark masses. We extrapolate these results both to the physical $u/d$
quark mass and to the physical $s$ quark mass. In section~\ref{SEC576}
we make predictions regarding the spectrum of $B$ and $B_s$ mesons by
interpolating  in the heavy quark mass to the physical $b$ quark mass
using experimental results as input. We close with a summary and a brief
outlook (section~\ref{SEC488}).


\section{\label{SEC466}Twisted mass lattice QCD}


\subsection{Simulation details}

We use $L^3 \times T = 24^3 \times 48$ gauge configurations produced by
the European Twisted Mass Collaboration (ETMC). The gauge action is the
tree-level Symanzik (tlSym) action \cite{Weisz:1982zw}
\begin{eqnarray}
S_\mathrm{G}[U] \ \ = \ \ \frac{\beta}{6} \bigg(b_0 \sum_{x,\mu\neq\nu} \textrm{Tr}\Big(1 - P^{1 \times 1}(x;\mu,\nu)\Big) + 
b_1 \sum_{x,\mu\neq\nu} \textrm{Tr}\Big(1 - P^{1 \times 2}(x;\mu,\nu)\Big)\bigg)
\end{eqnarray}
 with the normalisation condition $b_0 = 1 - 8 b_1$ and $b_1 = -1/12$. 
The fermionic action is the Wilson twisted mass (Wtm) action
\cite{Frezzotti:2000nk,Frezzotti:2003ni,Shindler:2007vp} with $N_f = 2$
degenerate flavours
\begin{eqnarray}
\label{eq:WtmQCD} S_\mathrm{F}[\chi,\bar{\chi},U] \ \ = \ a^4 \sum_x \bar{\chi}(x) \Big(D_{\rm W} + i\mu_\mathrm{q}\gamma_5\tau_3\Big) \chi(x) , 
\end{eqnarray}
where
\begin{eqnarray}
\label{eq:Wilson} D_\mathrm{W} \ \ = \ \ \frac{1}{2} \Big(\gamma_\mu \Big(\nabla_\mu + \nabla^\ast_\mu\Big) - a \nabla^\ast_\mu \nabla_\mu\Big) + m_0 ,
\end{eqnarray}
 $\nabla_\mu$ and $\nabla^\ast_\mu$ are the standard gauge covariant
forward and backward derivatives, $m_0$ and $\mu_\mathrm{q}$ are the
bare untwisted and twisted quark masses respectively and $\chi =
(\chi^{(u)} \, , \, \chi^{(d)})$ represents the fermionic field in the
so-called twisted basis. It is useful to introduce at this point the
twist angle $\omega$ given by $\tan \omega = \mu_\mathrm{R} /
m_\mathrm{R}$, where $\mu_\mathrm{R}$ and $m_\mathrm{R}$ denote the
renormalised twisted and untwisted quark masses. This angle
characterises the particular lattice action and must be kept fixed up to
$\mathcal{O}(a)$, while performing the continuum limit.

The results presented in this paper have been obtained with gauge configurations computed at $\beta = 3.9$ corresponding to a lattice spacing $a = 0.0855(5) \, \textrm{fm}$. We consider five different values of $\mu_\mathrm{q}$ with $m_0$ tuned to its critical value at $\mu_\mathrm{q} = 0.0040$ \cite{Boucaud:2007uk,Urbach:2007rt,Boucaud:2008xu} (cf.\ Table~\ref{TAB003}, where for each value the corresponding ``pion mass'' $m_\mathrm{PS}$ and number of gauge configurations is listed). With this tuning our target continuum theory is given by
\begin{eqnarray}
\mathcal{L} \ \ = \ \ \bar{\chi}(x) \Big(\gamma_\mu D_\mu + i \mu_\mathrm{R} \gamma_5 \tau_3\Big) \chi(x),
\end{eqnarray}
 which is parameterised by the renormalised twisted quark mass
$\mu_\mathrm{R}$. The tuning guarantees automatic $\mathcal{O}(a)$
improvement for physical correlation functions involving only light
fermions \cite{Frezzotti:2003ni}. In section~\ref{SEC496} we will argue
that automatic $\mathcal{O}(a)$ improvement also holds for static-light
spectral quantities without additional complications.



\begin{table}[h!]
\begin{center}
\begin{tabular}{|c|c|c|}
\hline
 & & \vspace{-0.40cm} \\
$\mu_\mathrm{q}$ & $m_\mathrm{PS}$ in MeV & number of gauge configurations \\
 & & \vspace{-0.40cm} \\
\hline
 & & \vspace{-0.40cm} \\
$0.0040$ & $314(2)$ & $1400$ \\
$0.0064$ & $391(1)$ & $1450$ \\
$0.0085$ & $448(1)$ & $1350$ \\
$0.0100$ & $485(1)$ & $900$ \\
$0.0150$ & $597(2)$ & $1000$\vspace{-0.40cm} \\
 & & \\
\hline
\end{tabular}
 \caption{\label{TAB003}bare twisted quark masses $\mu_\mathrm{q}$, pion
masses $m_\mathrm{PS}$ and number of gauge configurations.}
 \end{center}
\end{table}


\subsection{\label{SEC489}Static-light correlation functions}

 To compute correctly  a static-light correlation function with the Wtm
lattice action (\ref{eq:WtmQCD}), we follow the general procedure
described in \cite{Frezzotti:2000nk} and reviewed in
\cite{Shindler:2007vp}. The procedure reads:
\begin{itemize}
 \item[(1)] start with the continuum static-light correlation function
you are interested in,

\item[(2)] perform the axial rotation 
\begin{eqnarray}
\label{eq:axial} \psi \ \  = \ \ \exp\Big(i \omega \gamma_5 \tau_3 / 2\Big) \chi \quad , \quad \bar{\psi} \ \ = \ \ \bar{\chi} \exp\Big(i \omega \gamma_5 \tau_3 / 2\Big)
\end{eqnarray}
 on the fields appearing in the correlation function with a given value
for $\omega$,

\item[(3)] compute the resulting correlation function with the Wtm
lattice action (\ref{eq:WtmQCD}), with a choice of quark masses, such
that $\tan \omega = \mu_\mathrm{R} / m_\mathrm{R}$ up to
$\mathcal{O}(a)$,

\item[(4)] perform the continuum limit with renormalisation constants
computed in a massless scheme, tuning the untwisted bare quark mass in
order to achieve the desired target continuum theory, i.e.\ the desired
value of the twist angle $\omega$.
 \end{itemize}
 Each value of $\omega$ defines a different discretisation, but when the
continuum limit is performed the result will be exactly the initially
chosen static-light correlation function in the continuum with quark
mass  $M_\mathrm{R}^2 = m_\mathrm{R}^2 + \mu_\mathrm{R}^2$.

 In the following we give an explicit example. In QCD the pseudoscalar
and scalar static-light currents read
\begin{eqnarray}
\mathcal{P}^\textrm{stat}(x) \ \ = \ \ \bar{Q}(x) \gamma_5 \psi^{(u)}(x) \quad , \quad \mathcal{S}^\textrm{stat}(x) \ \ = \ \ \bar{Q}(x) \psi^{(u)}(x) ,
\end{eqnarray}
 where $Q$ is the static quark field\footnote{We will discuss the static
quark action in section~\ref{SEC639}.} and $\psi^{(u)}$ is a single
flavour of the light fermion doublet \\ $\psi = (\psi^{(u)} \, , \,
\psi^{(d)})$. Let us suppose we are interested in computing in continuum
QCD the static-light pseudoscalar-pseudoscalar correlation function
\begin{eqnarray}
\label{eq:AA} \mathcal{C}_\mathcal{PP} \ \ = \ \ \Big\langle (\mathcal{P}^\textrm{stat})_{\mathrm{R}}(x) (\mathcal{P}^\textrm{stat})^{\dagger}_\mathrm{R}(y) \Big\rangle_{(M_\mathrm{R},0)} ,
\end{eqnarray}
 where we write an index $(M_\mathrm{R},0)$ to specify that the
continuum action has a vanishing twisted mass and a renormalised
untwisted mass given by $M_{\rm R}$. We perform the axial rotation
(\ref{eq:axial}) obtaining
\begin{eqnarray}
\label{eq:corr_tw} \cos^2(\omega / 2) Z_\mathrm{P}^2 C_\mathrm{PP} + \sin^2(\omega / 2) Z_\mathrm{S}^2 C_\mathrm{SS} - i \cos(\omega / 2) \sin(\omega / 2) Z_\mathrm{P} Z_\mathrm{S} \Big(C_\mathrm{PS} - C_\mathrm{SP}\Big) ,
\end{eqnarray}
 where $Z_\mathrm{P}$ and $Z_\mathrm{S}$ are the standard
renormalisation constants for static-light currents computed in a
massless scheme with Wilson fermions. Note that for the static-light case,
 $Z_\mathrm{V} \equiv Z_\mathrm{P}$ and $Z_\mathrm{A} \equiv
Z_\mathrm{S}$. This correlation function has to be computed with the Wtm
action (\ref{eq:WtmQCD}) with quark masses tuned accordingly to the
value of $\omega$ chosen. The $C_\mathrm{XX}$ correlation functions in
(\ref{eq:corr_tw}) are defined in terms of currents in the twisted basis
\begin{eqnarray}
C_\mathrm{PP} \ \ = \ \ \Big\langle P^\textrm{stat}(x) (P^\textrm{stat})^\dagger(y) \Big\rangle_{(m_\mathrm{R},\mu_\mathrm{R})} \quad , \quad C_\mathrm{SS} \ \ = \ \ \Big\langle S^\textrm{stat}(x) (S^\textrm{stat})^\dagger(y) \Big\rangle_{(m_\mathrm{R},\mu_\mathrm{R})} \quad , \quad \ldots ,
\end{eqnarray}
where
\begin{eqnarray}
P^\textrm{stat}(x) \ \ = \ \ \bar{Q} \gamma_5 \chi^{(u)}(x) \quad , \quad S^\textrm{stat}(x) \ \ = \ \ \bar{Q}(x) \chi^{(u)}(x) .
\end{eqnarray}
 Once the continuum limit of the correlation function (\ref{eq:corr_tw})
has been performed, the result will be the original correlation function
(\ref{eq:AA}) with $M_\mathrm{R}^2 = m_\mathrm{R}^2 + \mu_\mathrm{R}^2$.

However, to compute spectral quantities it is sufficient to analyze a
matrix of correlation functions of bare currents with the appropriate
quantum numbers. We will discuss this in detail in
section~\ref{sec:parity_mixing}.


\subsection{\label{SEC496}Automatic $\mathcal{O}(a)$ improvement of
static-light meson masses}

Spectral quantities like hadron masses extracted from lattice
simulations of  Wilson fermions will in general be affected by
$\mathcal{O}(a)$ discretisation errors.  In the particular case of
masses extracted from static-light correlation functions the
$\mathcal{O}(a)$  discretisation errors come from the
dimension-5-operators of the Symanzik effective action of the light and
static quarks.

The Symanzik effective action for the Eichten-Hill (EH) static action
contains only one term, which contributes to the $\mathcal{O}(a)$
corrections of the linearly divergent static self-energy
\cite{Kurth:2000ki}. In this paper all observables we consider are
differences, where this static self-energy cancels. Moreover, this
result is independent on the particular lattice static action chosen, as
long as it preserves the relevant symmetries of the EH action. This is
the case for our choice of static action (cf.\ section~\ref{SEC639}).

As a consequence, the only $\mathcal{O}(a)$ errors which could affect
our results, come from the dimension-5-operators of the Symanzik
effective action of the light quarks. The light quark action used in
this paper is Wtm at maximal twist. It is by now well known that at
maximal twist a single insertion of a dimension-5-operator of the
Symanzik effective action into parity even correlation functions
vanishes, because, independently on the lattice basis adopted, these
operators are parity odd and the insertions have to be evaluated in the
continuum theory, where parity is a preserved symmetry
\cite{Frezzotti:2003ni}. We can conclude that all the spectral
quantities, when the static self-energy has been removed, are
automatically $\mathcal{O}(a)$ improved.


\subsection{\label{sec:parity_mixing}Spectral decomposition and parity mixing}

In this section we explain, how to analyze lattice results for
static-light correlation functions obtained in the twisted basis. In
particular we concentrate on the assignment of parity labels to
extracted static-light meson states.

We start from the physical basis and, for simplicity, consider only two
operators, the pseudoscalar and the scalar static-light current, and
only two states, which we label by $| 1 \rangle$ and $| 2 \rangle$. The
explanation carries over to the more general case in a straightforward
way.

First consider the following matrix of correlation functions in the physical basis:
\begin{eqnarray}
\mathcal{C}(t) \ \ = \ \ \left(\begin{array}{cc}
\mathcal{C}_\mathcal{PP}(t) & \mathcal{C}_\mathcal{PS}(t) \\
\mathcal{C}_\mathcal{SP}(t) & \mathcal{C}_\mathcal{SS}(t)
\end{array}\right) ,
\end{eqnarray}
 where $\mathcal{C}_\mathcal{PP}(t)$ has been defined in (\ref{eq:AA})
with $x = (t,\vec{0})$ and $y = (0,\vec{0})$ and analogously the others.
The parity of the operators $(\mathcal{P}^\textrm{stat})_{\mathrm{R}}$
and $(\mathcal{S}^\textrm{stat})_{\mathrm{R}}$ is determined by the
parity transformation properties of the associated field, i.e.\
$(\mathcal{P}^\textrm{stat})_{\mathrm{R}}$ has negative parity and
$(\mathcal{S}^\textrm{stat})_{\mathrm{R}}$ has positive parity. Even if
parity is broken at finite lattice spacing, one can still assign a
parity label to each of the states we use to decompose the correlation
functions \cite{Frezzotti:2003ni}. If we consider only two states, the
spectral decomposition will have the form
\begin{eqnarray}
\label{EQN954} \mathcal{C}(T) \ \ = \ \  \left(\begin{array}{cc}
|a_1^\mathcal{P}|^2 & (a_1^\mathcal{P})^\ast a_1^\mathcal{S} \\ 
(a_1^\mathcal{S})^\ast a_1^\mathcal{P} & |a_1^\mathcal{S}|^2
\end{array}\right) e^{-M_1 t} +  \left(\begin{array}{cc}
|a_2^\mathcal{P}|^2 & (a_2^\mathcal{P})^\ast a_2^\mathcal{S} \\ 
(a_2^\mathcal{S})^\ast a_2^\mathcal{P} & |a_2^\mathcal{S}|^2
\end{array}\right) e^{-M_2 t} ,
\end{eqnarray}
where we have defined
\begin{eqnarray}
(a_{1,2}^\mathcal{P})^\ast \ \ = \ \ \langle \Omega | 
\hat{\mathcal{P}}^\textrm{stat} | 1,2 \rangle \quad , 
\quad (a_{1,2}^\mathcal{S})^\ast \ \ = \ \ \langle \Omega | \hat{\mathcal{S}}^\textrm{stat} | 1,2 \rangle .
\end{eqnarray}
The correlation functions $\mathcal{C}_\mathcal{PS}$ and
$\mathcal{C}_\mathcal{SP}$ vanish in the continuum limit, because parity
is a symmetry of QCD. This means by universality that at finite lattice
spacing they are at most of $\mathcal{O}(a)$. Since
$\mathcal{C}_\mathcal{PP}$ and $\mathcal{C}_\mathcal{SS}$ are of
$\mathcal{O}(1)$ in the continuum limit, we can conclude that for given
$n$ either $a_n^\mathcal{P}$ is of $\mathcal{O}(1)$ and
$a_n^\mathcal{S}$ is of $\mathcal{O}(a)$ or the opposite way round
\cite{Frezzotti:2003ni}. We can conclude that if $a_n^\mathcal{P}$ is of
$\mathcal{O}(1)$, the state $| n \rangle$ has the same parity as the
formal parity of $\mathcal{P}^\textrm{stat}$, which in this case is
negative. Moreover, $a_n^\mathcal{S}$ is of $\mathcal{O}(a)$ and has
to vanish in the continuum limit.

 We now perform the axial transformation (\ref{eq:axial}). The relation
between correlation functions up to discretisation errors is, for
example, for $\mathcal{C}_\mathcal{PP}$
\begin{eqnarray}
\label{EQN955} \mathcal{C}_\mathcal{PP} \ \ = \ \ \cos^2(\omega / 2) Z_\mathrm{P}^2 C_\mathrm{PP} + \sin^2(\omega / 2) Z_\mathrm{S}^2 C_\mathrm{SS} - i \cos(\omega / 2) \sin(\omega / 2) Z_\mathrm{P} Z_\mathrm{S} \Big(C_\mathrm{PS} - C_\mathrm{SP}\Big) .
\end{eqnarray}
For the matrix of correlation functions in the twisted basis
\begin{eqnarray}
C(t) \ \ = \ \ \left(\begin{array}{cc}
C_\mathrm{PP}(t) & C_\mathrm{PS}(t) \\
C_\mathrm{SP}(t) & C_\mathrm{SS}(t)
\end{array}\right)
\end{eqnarray}
 we can also perform a spectral decomposition considering again only the
states $| 1 \rangle$ and $| 2 \rangle$:
\begin{eqnarray}
\label{EQN956} C(t) \ \ = \ \ \left(\begin{array}{cc}
|b_1^\mathrm{P}|^2 & (b_1^\mathrm{P})^\ast b_1^\mathrm{S} \\ 
(b_1^\mathrm{S})^\ast b_1^\mathrm{P} & |b_1^\mathrm{S}|^2
\end{array}\right) e^{-M_1 t} + \left(\begin{array}{cc}
|b_2^\mathrm{P}|^2 & (b_2^\mathrm{P})^\ast b_2^\mathrm{S} \\ 
(b_2^\mathrm{S})^\ast b_2^\mathrm{P} & |b_2^\mathrm{S}|^2
\end{array}\right) e^{-M_2 t} .
\end{eqnarray}
From (\ref{EQN954}), (\ref{EQN955}) and (\ref{EQN956}) we can conclude
\begin{eqnarray}
\nonumber & & \hspace{-0.7cm} |a_{1,2}^\mathcal{P}|^2 \ \ = \ \ \cos^2(\omega / 2) Z_\mathrm{P}^2 |b_{1,2}^\mathrm{P}|^2 + \sin^2(\omega / 2) Z_\mathrm{S}^2 |b_{1,2}^\mathrm{S}|^2 + 2 \cos(\omega / 2) \sin(\omega / 2) Z_\mathrm{P} Z_\mathrm{S} \mathrm{Im}\Big((b_{1,2}^\mathrm{P})^\ast b_{1,2}^\mathrm{S}\Big) \\
\label{eq:ampl_rel} & & \\
\nonumber & & \hspace{-0.7cm} |a_{1,2}^\mathcal{S}|^2 \ \ = \ \ \cos^2(\omega / 2) Z_\mathrm{S}^2 |b_{1,2}^\mathrm{S}|^2 + \sin^2(\omega / 2) Z_\mathrm{P}^2 |b_{1,2}^\mathrm{P}|^2 + 2 \cos(\omega / 2) \sin(\omega / 2) Z_\mathrm{P} Z_\mathrm{S} \mathrm{Im}\Big((b_{1,2}^\mathrm{S})^\ast b_{1,2}^\mathrm{P}\Big) . \\
 & & 
\end{eqnarray}
 If the state $| 1 \rangle$ has negative parity, $|a_1^\mathcal{S}|^2$
has to vanish as $\mathcal{O}(a^2)$ in the continuum limit, while
$|a_1^\mathcal{P}|^2$ has to be of $\mathcal{O}(1)$. Since the first two
terms on the right hand side of (\ref{eq:ampl_rel}) are positive and
non-vanishing in the continuum limit,  there must be a cancellation
coming from the third term. In fact we immediately see that this third
term has opposite sign for $|a_{1,2}^\mathcal{P}|^2$ compared to
$|a_{1,2}^\mathcal{S}|^2$. This allows us to identify the parity of the
states $|1 \rangle$ and $|2 \rangle$ without knowing the exact values of
the renormalisation constants and the twist angle. The criterion will be
the following: if
\begin{eqnarray}
\mathrm{Im}\Big((b_{1}^\mathrm{S})^\ast b_{1}^\mathrm{P}\Big) \ \ < \ \ 0 ,
\end{eqnarray}
 the state $|1 \rangle$ has negative parity, otherwise positive parity.
The other cases follow accordingly.

This method, which we have described for a simple case, is valid
independently of the number of states considered and the kind of
operators studied. At finite lattice spacing it provides a way to assign
a formal parity to each of the extracted states.

The method extends to all cases, where the light degrees of freedom
involve fermions in the twisted basis, e.g.\ for static-light mesons,
but also for baryons.


\section{\label{SEC399}The static-light meson spectrum}


\subsection{Static-light trial states}


\subsubsection{\label{SEC052}Static-light meson creation operators in
the continuum}

 It is convenient to discuss static-light mesons treating the static
quark as a four component spinor since the symmetries of hadronic
bilinears are well studied \cite{Lacock:1996vy}. In the continuum an
operator creating a static-light meson with well defined quantum numbers
$J$ , $j$ and $\mathcal{P}$ is given by
\begin{eqnarray}
\label{EQN001} \mathcal{O}^{(\Gamma)}(\mathbf{x}) \ \ = \ \ \bar{Q}(\mathbf{x}) \int d\hat{\mathbf{n}} \, \Gamma(\hat{\mathbf{n}}) U(\mathbf{x};\mathbf{x}+r \hat{\mathbf{n}}) \psi^{(u)}(\mathbf{x}+r \hat{\mathbf{n}}) .
\end{eqnarray}
$\bar{Q}(\mathbf{x})$ represents an infinitely heavy antiquark (here a Dirac
spinor) at position $\mathbf{x}$, $\int d\hat{\mathbf{n}}$ denotes an
integration over the unit sphere, $U$ is a straight parallel transporter
and $\psi^{(u)}(\mathbf{x}+r \hat{\mathbf{n}})$ creates a light quark at
position $\mathbf{x}+r \hat{\mathbf{n}}$ separated by a distance $r$
from the antiquark (of course, using $\psi^{(d)}$ instead of
$\psi^{(u)}$ would yield identical results). $\Gamma$ is an appropriate
combination of spherical harmonics and $\gamma$ matrices coupling
angular momentum and quark spin to yield well defined total angular
momentum $J$ (static quark spin included) and $j$ (static quark spin not
included) and parity $\mathcal{P}$. The meson creation operators used in
the following are listed in Table~\ref{TAB001}.

\begin{table}[h!]
\begin{center}

\begin{tabular}{|c||c|c||c|c||c|}
\hline
 & & & & & \vspace{-0.40cm} \\
$\Gamma(\hat{\mathbf{n}})$ &  $J^\mathcal{P}$ & $j^\mathcal{P}$ & $\mathrm{O}_\mathrm{h}$ & lattice $j^\mathcal{P}$ & notation \\
 & & & & & \vspace{-0.40cm} \\
\hline
 & & & & & \vspace{-0.41cm} \\
\hline
 & & & & & \vspace{-0.40cm} \\
$\gamma_5 \ , \ \gamma_5 \gamma_j \hat{n}_j$ & $0^- \ [1^-]$ & $(1/2)^-$ & $A_1$ & $(1/2)^- \ , \ (7/2)^- \ , \ ...$ & $S$ \\
 & & & & & \vspace{-0.40cm} \\
$1 \ , \ \gamma_j \hat{n}_j$ &  $0^+ \ [1^+]$ & $(1/2)^+$ & & $(1/2)^+ \ , \ (7/2)^+ \ , \ ...$ & $P_-$ \\
 & & & & & \vspace{-0.40cm} \\
\hline
 & & & & & \vspace{-0.40cm} \\
$\gamma_1 \hat{n}_1 - \gamma_2 \hat{n}_2$ (and cyclic) & $2^+ \ [1^+]$ & $(3/2)^+$ & $E$ & $(3/2)^+ \ , \ (5/2)^+ \ , \ ...$ & $P_+$ \\
 & & & & & \vspace{-0.40cm} \\
$\gamma_5 (\gamma_1 \hat{n}_1 - \gamma_2 \hat{n}_2)$ (and cyclic) & $2^- \ [1^-]$ & $(3/2)^-$ & & $(3/2)^- \ , \ (5/2)^- \ , \ ...$ & $D_\pm$ \\
 & & & & & \vspace{-0.40cm} \\
\hline
 & & & & & \vspace{-0.40cm} \\
$\gamma_1 \hat{n}_2 \hat{n}_3 + \gamma_2 \hat{n}_3 \hat{n}_1 + \gamma_3 \hat{n}_1 \hat{n}_2$ & $ 3^- \ [2^-]$ & $(5/2)^-$ & $A_2$ & $(5/2)^- \ , \ (7/2)^- \ , \ ...$ & $D_+$ \\
 & & & & & \vspace{-0.40cm} \\
$\gamma_5 (\gamma_1 \hat{n}_2 \hat{n}_3 + \gamma_2 \hat{n}_3 \hat{n}_1 + \gamma_3 \hat{n}_1 \hat{n}_2)$ & $3^+ \ [2^+]$ & $(5/2)^+$ & & $(5/2)^+ \ , \ (7/2)^+ \ , \ ...$ & $F_\pm$\vspace{-0.40cm} \\
 & & & & & \\
\hline
\end{tabular}

 \caption{\label{TAB001}Static-light meson creation operators. The other
mesonic $J^\mathcal{P}$ states that are degenerate with that created are
noted in square brackets.}
 \end{center}
\end{table}


\subsubsection{\label{SEC053}Static-light meson creation operators on a lattice}

 Here we present the construction of appropriate lattice operators to
create the states of interest, following
\cite{Michael:1998sg,Lacock:1996vy}. When putting static-light meson
creation operators (\ref{EQN001}) on a lattice, one has to replace the
integration over the unit sphere by a discrete sum over lattice sites,
which have the same distance from the static antiquark at position
$\mathbf{x}$. For the operators in $A_1$ and $E$ representations we use
six lattice sites, i.e.\
\begin{eqnarray}
\label{EQN002} \mathcal{O}^{(\Gamma)}(\mathbf{x}) \ \ = \ \ \bar{Q}(\mathbf{x}) \sum_{\mathbf{n} = \pm \hat{\mathbf{e}}_1 , \pm \hat{\mathbf{e}}_2 , \pm \hat{\mathbf{e}}_3} \Gamma(\hat{\mathbf{n}}) U(\mathbf{x};\mathbf{x}+r \mathbf{n}) \chi^{(u)}(\mathbf{x}+r \mathbf{n}) ,
 \end{eqnarray}
 whereas for those in the $A_2$ representation one has to use eight
lattice sites, i.e.\
\begin{eqnarray}
\label{EQN003} \mathcal{O}^{(\Gamma)}(\mathbf{x}) \ \ = \ \ \bar{Q}(\mathbf{x}) \sum_{\mathbf{n} = \pm \hat{\mathbf{e}}_1 \pm \hat{\mathbf{e}}_2 \pm \hat{\mathbf{e}}_3} \Gamma(\hat{\mathbf{n}}) U(\mathbf{x};\mathbf{x}+r \mathbf{n}) \chi^{(u)}(\mathbf{x}+r \mathbf{n}) .
\end{eqnarray} 
 In the first case the spatial parallel transporters are straight paths
of links, while in the second case we use ``diagonal links'', which are
averages over the six possible paths around a cube between opposite
corners projected back to SU(3).

The states created by these lattice meson creation operators do not form
irreducible representations of the rotation group $\mathrm{SO}(3)$, but
of its cubic subgroup $\mathrm{O}_\mathrm{h}$. Therefore, these states
have no well defined total angular momentum, but are linear
superpositions of an infinite number of total angular momentum
eigenstates. The common notation of the corresponding
$\mathrm{O}_\mathrm{h}$ representations together with their lowest
angular momentum content are also listed in Table~\ref{TAB001}. Note
that we do not consider $\mathrm{O}_\mathrm{h}$ representations $T_1$
and $T_2$, because these representations yield correlation functions,
which are numerically identical to those listed (e.g.\ $T_1$ would be
$\Gamma = \gamma_j$ or $\Gamma = \gamma_5 \gamma_j$, which gives the
same correlations as $\Gamma = \gamma_5$ and $\Gamma = 1$, and $T_2$
would be $\Gamma = \gamma_1 n_2 + \gamma_2 n_1$ or $\Gamma = \gamma_5
(\gamma_1 n_2 + \gamma_2 n_1)$, which gives the same correlations as
$\Gamma = \gamma_1 n_1 - \gamma_2 n_2$ and $\Gamma = \gamma_5 (\gamma_1
n_1 - \gamma_2 n_2)$).

Since the $D_-$ and the $D_+$ states as well as the $F_-$ and $F_+$
states are expected to have a similar mass, we do not have unambiguous
lattice operators to determine $D_-$ and $F_-$ but rather operators, which have
an admixture of $D_{\pm}$ and $F_{\pm}$ respectively. We label these
operators as $D_{\pm}$ and $F_{\pm}$ (cf.\ Table~\ref{TAB001}).

We have also replaced the light quark fields in the physical basis
$\psi^{(u)}$ by their counterparts in the twisted basis $\chi^{(u)}$.
Note that trial states created by such twisted basis operators are not
eigenstates of parity. Nevertheless, as we have discussed in
section~\ref{sec:parity_mixing}, it is possible to assign unambiguously
a parity label to the masses extracted from the time dependence of such twisted basis correlators.


\subsubsection{Smearing techniques}

When performing a lattice study of the static-light meson spectrum, the
following points have to be considered:
\begin{itemize}
 \item It is imperative to use trial states with large overlap to low
lying energy eigenstates.  Only then the corresponding meson masses can
be extracted from correlation functions at small temporal separations, 
where signal-to-noise ratios are acceptable.

 \item To determine excited states for a given $\mathrm{O}_\mathrm{h}$
representation,  it is necessary to have a whole set of linearly
independent trial states belonging to that $\mathrm{O}_\mathrm{h}$
representation.
 \end{itemize}
 To fulfill both requirements we use different ``radii'' $r$ (cf.\
eqns.\ (\ref{EQN002}) and (\ref{EQN003}))  and apply APE smearing and
Gaussian smearing also with different parameters.
 The resulting extended trial states have significantly better  overlap
to low lying energy eigenstates than their unsmeared counterparts.

\subsubsection*{APE smearing of spatial links}

After $N_\mathrm{APE}$ iterations APE smeared spatial links
\cite{Albanese:1987ds} are given by
 \begin{eqnarray}
\nonumber & & \hspace{-0.7cm} U^{(N_\mathrm{APE})}(x,x+e_k) \ \ = \ \ P_\mathrm{SU(3)}\bigg(U^{(N_\mathrm{APE}-1)}(x,x+e_k) + 
\alpha_\mathrm{APE} \sum_{j = \pm 1,\pm 2,\pm 3}^{j \neq \pm k} U^{(N_\mathrm{APE}-1)}(x,x+e_j) \\
 & & \hspace{0.675cm} U^{(N_\mathrm{APE}-1)}(x+e_j,x+e_j+e_k) U^{(N_\mathrm{APE}-1)}(x+e_j+e_k,x+e_k)\bigg) ,
\end{eqnarray}
 where $U^{(0)}$ are the original unsmeared links. $\alpha_\mathrm{APE}$
is a weight parameter and $P_\mathrm{SU(3)}$ denotes a projection back
to $\mathrm{SU(3)}$ defined by
 \begin{eqnarray}
\label{EQN943} P_\mathrm{SU(3)}(U) \ \ = \ \ \frac{U'}{\det(U')^{1/3}} \quad , \quad U' \ \ = \ \ U \Big(U^\dagger U\Big)^{-1/2}
 \end{eqnarray}
 with $\det(U')^{1/3}$ being that root closest to $1$.

\subsubsection*{Gaussian smearing of light quark operators}

 After $N_\textrm{Gauss}$ iterations Gaussian smeared light quark
operators \cite{Gusken:1989qx,Alexandrou:2008tn} are given by
\begin{eqnarray}
\nonumber & & \hspace{-0.7cm} \chi^{(N_\textrm{Gauss})}(x) \ \ = \\
\label{EQN586} & & = \ \ \frac{1}{1 + 6 \kappa} \bigg(\chi^{(N_\textrm{Gauss}-1)}(x) + \kappa_\textrm{Gauss} \sum_{j=\pm 1,\pm 2,\pm 3} 
U^{(N_\mathrm{APE})}(x,x+e_j) \chi^{(N_\textrm{Gauss}-1)}(x+e_j)\bigg) ,
\end{eqnarray}
 where $\chi^{(0)}$ are the original unsmeared light quark operators and
$U^{(N_\mathrm{APE})}$ denote APE smeared spatial links.


\subsection{Correlation matrices}

 For each $\mathrm{O}_\mathrm{h}$ representation we compute  $6 \times
6$ correlation matrices
 \begin{eqnarray}
\label{EQN201} C_{K K'}(t) \ \ = \ \ \Big\langle \mathcal{O}^{(K)}(t) (\mathcal{O}^{(K')})^\dagger(0) \Big\rangle ,
 \end{eqnarray}
 where $\mathcal{O}^{(K)}$ is a static-light meson creation operator
(cf.\ eqns.\ (\ref{EQN002}) and (\ref{EQN003}))  with $K$ denoting its
parameters, i.e.\ $K = (\Gamma \, , \, N_\textrm{Gauss} \, , \, r)$  (we
have chosen $N_\mathrm{APE} = 10$, $\alpha_\mathrm{APE} = 0.5$ and
$\kappa_\textrm{Gauss} = 0.5$ for all operators).  Detailed information
about the operator content of the correlation matrices is given in
Table~\ref{TAB002}.

\begin{table}[h!]
\begin{center}

\begin{tabular}{|c|c|c|c|c|c|}
\hline
 & & & & & \vspace{-0.40cm} \\
$\mathrm{O}_\mathrm{h}$ & $\Gamma$ & $N_\textrm{Gauss}$ & $r$ & $R/a$ & $R$ in fm \\
 & & & & & \vspace{-0.40cm} \\
\hline
 & & & & & \vspace{-0.41cm} \\
\hline
 & & & & & \vspace{-0.40cm} \\
$A_1$ & $\gamma_5$ & $30$ & $3$ & $5.61$ & $0.48$ \\
      &            & $60$ & $6$ & $9.00$ & $0.77$ \\
 & & & & & \vspace{-0.40cm} \\
\cline{2-6}
 & & & & & \vspace{-0.40cm} \\
 & $1$ & $30$ & $3$ & $5.61$ & $0.48$ \\
 &     & $60$ & $6$ & $9.00$ & $0.77$ \\
 & & & & & \vspace{-0.40cm} \\
\cline{2-6}
 & & & & & \vspace{-0.40cm} \\
 & $\gamma_5 \gamma_j x_j$ & $30$ & $3$ & $5.61$ & $0.48$ \\
 & & & & & \vspace{-0.40cm} \\
\cline{2-6}
 & & & & & \vspace{-0.40cm} \\
 & $\gamma_j x_j$ & $30$ & $3$ & $5.61$ & $0.48$ \\
 & & & & & \vspace{-0.40cm} \\
\hline
 & & & & & \vspace{-0.41cm} \\
\hline
 & & & & & \vspace{-0.40cm} \\
$E$ & $\gamma_1 x_1 - \gamma_2 x_2$ (and cyclic) & $30$ & $3$ & $5.61$ & $0.48$ \\
    &                                            & $60$ & $6$ & $9.00$ & $0.77$ \\
    &                                            & $90$ & $3$ & $8.74$ & $0.75$ \\
 & & & & & \vspace{-0.40cm} \\
\cline{2-6}
 & & & & & \vspace{-0.40cm} \\
 & $\gamma_5 (\gamma_1 x_1 - \gamma_2 x_2)$ (and cyclic) & $30$ & $3$ & $5.61$ & $0.48$ \\
 &                                                       & $60$ & $6$ & $9.00$ & $0.77$ \\
 &                                                       & $90$ & $3$ & $8.74$ & $0.75$ \\
 & & & & & \vspace{-0.40cm} \\
\hline
 & & & & & \vspace{-0.41cm} \\
\hline
 & & & & & \vspace{-0.40cm} \\
$A_2$ & $\gamma_1 x_2 x_3 + \gamma_2 x_3 x_1 + \gamma_3 x_1 x_2$ & $30$ & $2$ & $5.88$ & $0.50$ \\
      &                                                          & $60$ & $4$ & $9.64$ & $0.82$ \\
      &                                                          & $90$ & $2$ & $8.91$ & $0.76$ \\
 & & & & & \vspace{-0.40cm} \\
\cline{2-6}
 & & & & & \vspace{-0.40cm} \\
 & $\gamma_5 (\gamma_1 x_2 x_3 + \gamma_2 x_3 x_1 + \gamma_3 x_1 x_2)$ & $30$ & $2$ & $5.88$ & $0.50$ \\
 &                                                                     & $60$ & $4$ & $9.64$ & $0.82$ \\
 &                                                                     & $90$ & $2$ & $8.91$ & $0.76$\vspace{-0.40cm} \\
 & & & & & \\
\hline
\end{tabular}

 \caption{\label{TAB002}static-light meson creation operators used for
the $A_1$, $E$ and $A_2$ correlation matrices.}
 \end{center}
\end{table}

The width of a Gaussian smeared light quark operator (\ref{EQN586}) in
lattice units is approximately given by
\begin{eqnarray}
\sigma \ \ \approx \ \ \sqrt{\frac{2 N_\textrm{Gauss} \kappa_\textrm{Gauss}}{1 + 6 \kappa_\textrm{Gauss}}} .
 \end{eqnarray}
For $\kappa_\textrm{Gauss} = 0.5$ and $N_\textrm{Gauss} = (30 \, , \,
60 \, , \, 90)$ this amounts to $\sigma \approx (2.74 \, , \, 3.87 \, ,
\, 4.74)$. Taking also the parameter $r$ into account one can estimate
the radius of a static-light trial state: \\ $R/a = \sqrt{r^2 + 3
\sigma^2}$ for the $A_1$ and $E$ representations and $R/a = \sqrt{ 3 r^2 +
3 \sigma^2}$ for the $A_2$ representation. The radii of the trial states used
are also listed in Table~\ref{TAB002} both in lattice units and in
physical units.

Note that to identify the parity of states extracted via fitting it is
important to compute correlation matrices, which contain for each
operator $\Gamma$ also its counterpart $\gamma_5 \Gamma$ (cf.\
section~\ref{sec:parity_mixing}).


\subsubsection{\label{SEC639}Quark propagators}

When evaluating the correlations (\ref{EQN201}), both static quark
propagators and light quark propagators appear.  To improve
signal-to-noise ratios, we apply the following techniques.

\subsubsection*{Static quark propagators}

To improve the signal to noise ratio for
static-light correlation functions,  we use the HYP2 static action
\cite{Hasenfratz:2001hp,DellaMorte:2003mn,Della Morte:2005yc}.  Static
quark propagators are given by
\begin{eqnarray}
\Big\langle Q(x) \bar{Q}(y) \Big\rangle_{Q,\bar{Q}} \ \ = \ \ \delta^{(3)}(\mathbf{x}-\mathbf{y}) U^{(\textrm{HYP2})}(x;y) 
\bigg(\Theta(y_0-x_0) \frac{1 - \gamma_0}{2} + \Theta(x_0-y_0) \frac{1 + \gamma_0}{2}\bigg) ,
\end{eqnarray}
 where $\langle \ldots \rangle_{Q,\bar{Q}}$ denotes the integration over
the static quark fields and $U(x;y)$  is a path ordered product of HYP2
smeared links along the straight path from $x$ to $y$.

\subsubsection*{Light quark propagators}

To exploit translational invariance, it is imperative to use stochastic
methods for the light quark propagators.  The correlators can then be
evaluated at a large number of source points, while only a few
inversions of the lattice  Dirac operator have to be performed. One very
powerful method is maximal variance reduction \cite{Michael:1998sg}.  A
somewhat easier method to implement is to use stochastic sources on time
slices and this has been found  to give reasonable results
\cite{McNeile:2004rf}. Because we have inverted from such time-slice
sources  as part of our light-light meson studies
\cite{Boucaud:2007uk,Urbach:2007rt,Boucaud:2008xu},  we follow this
latter route, since it is computationally much quicker for us.

For each gauge configuration we use $N_s$ stochastic $\mathcal{Z}_2
\times \mathcal{Z}_2$ sources $\xi^{(\alpha)}$, $\alpha = 1,\ldots,N_s$
located on the same timeslice. For our lightest three $\mu_q$ values we
take $N_s = 4$ sources, which are the same for each of the four spin
components so that we can re-use previous inversions
\cite{Boucaud:2007uk,Urbach:2007rt,Boucaud:2008xu}. For our heavier two
$\mu_q$ values, we had to redo the inversions so we use only $N_s=1$
source with random values in each of the spin components.

After solving
\begin{eqnarray}
D_\mathrm{Wtm}^{(u)}(x;y) \phi^{(\alpha)}(y) \ \ = \ \ \xi^{(\alpha)}(x) ,
\end{eqnarray}
where $D_\mathrm{Wtm}^{(u)} = D_\mathrm{W} + i \mu_\mathrm{q} \gamma_5$ is the twisted mass Dirac operator acting on $\chi^{(u)}$, the light quark propagator is given by the unbiased estimate
\begin{eqnarray}
\Big\langle \chi^{(u)}(x) \bar{\chi}^{(u)}(y) \Big\rangle_{\chi,\bar{\chi}} \ \ = \ \ (D_\mathrm{Wtm}^{(u)})^{-1}(x;y) \ \ \approx \ \ \sum_{\alpha = 1}^{N_s} \phi^{(\alpha)}(x) (\xi^{(\alpha)})^\dagger(y) ,
\end{eqnarray}
 where $\langle \ldots \rangle_{\chi,\bar{\chi}}$ denotes the
integration over the light quark fields.


\subsection{\label{SEC729}Extracting static-light meson masses from correlation matrices}

 Assuming that for sufficiently large $t$ the correlation matrix
(\ref{EQN201}) can be approximated by the $n$ lowest lying energy
eigenstates $| i \rangle$, $i = 1,\ldots,n$ we use the ansatz
\begin{eqnarray}
\label{EQN638} \Big(\mathcal{O}^{(K)}\Big)^\dagger | \Omega \rangle \ \ \approx \ \ \sum_{i=1}^n b_i^K | i \rangle .
\end{eqnarray}
The correlation matrix (\ref{EQN201}) in terms of the ansatz is
\begin{eqnarray}
\label{EQN755} C_{K K'}(t) \ \ \approx \ \ \sum_{i=1}^n (b_i^K)^\ast b_i^{K'} e^{-E_i t} \ \ = \ \ \tilde{C}_{K K'}(t) .
\end{eqnarray}
The parameters $E_i$ and $b_i^K$ are determined by minimising
\begin{eqnarray}
\label{EQN756} \chi^2 \ \ = \ \ \sum_{t=t_\textrm{min}}^{t_\textrm{max}} \sum_{K \leq K'} \left(\frac{C_{K K'}(t) - \tilde{C}_{K K'}(t)}{\sigma(C_{K K'}(t))}\right)^2 ,
\end{eqnarray}
where $\sigma(C_{K K'}(t))$ denotes the statistical error of $C_{K K'}(t)$.

In the following we apply this fitting procedure with $n = 4$
exponentials. To obtain physically meaningful results with small
statistical errors, it is essential to determine an appropriate fitting
range $t_\textrm{min} \ldots t_\textrm{max}$. To this end, we have
performed correlated fits with various fitting ranges using eigenvalue
smoothed covariance matrices \cite{Michael:1994sz}. We have found that
$t_\textrm{min} = 3$ gives reasonable reduced $\chi^2$ values (cf.\
Table~\ref{TAB004}), while data points beyond $t_\textrm{max} = 12$ seem
to be dominated by statistical noise, i.e.\ including them in the fits
does not alter resulting meson masses nor corresponding statistical
errors.

\begin{table}[h!]
\begin{center}
\begin{tabular}{|c|c|c|c|c|c|}
\hline
 & & & & & \vspace{-0.40cm} \\
$\mathrm{O}_\mathrm{h}$ & $\mu_\mathrm{q} = 0.0040$ & $\mu_\mathrm{q} = 0.0064$ & $\mu_\mathrm{q} = 0.0085$ & $\mu_\mathrm{q} = 0.0100$ & $\mu_\mathrm{q} = 0.0150$ \\
 & & & & & \vspace{-0.40cm} \\
\hline
 & & & & & \vspace{-0.40cm} \\
$A_1$ & $1.89$ & $2.30$ & $2.35$ & $0.95$ & $1.16$ \\
$E$   & $1.21$ & $1.33$ & $1.70$ & $2.04$ & $2.09$ \\
$A_2$ & $1.56$ & $1.96$ & $1.28$ & $1.16$ & $1.26$\vspace{-0.40cm} \\
 & & & & & \\
\hline
\end{tabular}
\caption{\label{TAB004}$\chi^2 / \mathrm{dof}$ from correlated $\chi^2$ fits for different $\mathrm{O}_\mathrm{h}$ representations and different $\mu_\mathrm{q}$.}
\end{center}
\end{table}

As has already been discussed in section~\ref{SEC053}, it is difficult
to unambiguously determine the total angular momentum $j$ of a state
obtained from a lattice computation. This is, because for every
$\mathrm{O}_\mathrm{h}$ representation there exists an infinite number
of possible total angular momentum eigenstates (cf.\
Table~\ref{TAB001}). In the following, we assume that the low lying
states we are going to study have the lowest total angular momentum
possible, i.e.\ we assign $j = 1/2$ to states from $A_1$, $j = 3/2$ to
states from $E$ and $j = 5/2$ to states from $A_2$. Parity on the other hand
can directly be read off from the coefficients $b_i^K$ (cf.\
section~\ref{sec:parity_mixing}).

\begin{figure}[h!]
\begin{center}
\input{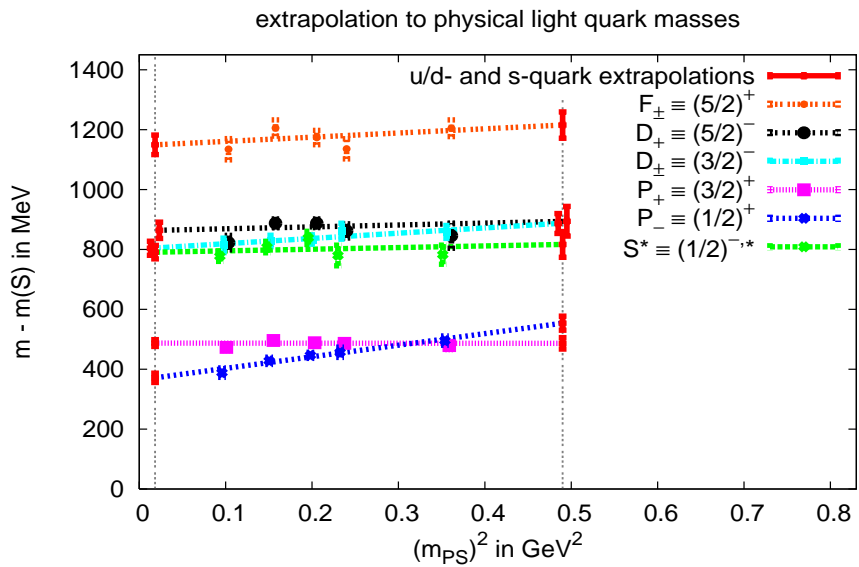}
\caption{\label{FIG004}static-light mass differences linearly extrapolated to the physical $u/d$ quark mass and the physical $s$ quark mass.}
\end{center}
\end{figure}

Since static-light meson masses diverge in the continuum limit due to
the self energy of the static quark, we always consider mass
differences, where this self energy cancels. Mass differences between
various static-light mesons with quantum numbers $j^\mathcal{P}$ and the
lightest static-light meson ($(1/2)^- \equiv S$ ground state) for all
five $\mu_\mathrm{q}$ values are collected in Figure~\ref{FIG004} and
Table~\ref{TAB005}. Statistical errors have been computed from $100$
bootstrap samples.

\begin{table}[h!]
\begin{center}
\begin{tabular}{|c|c|c|c|c|c|}
\hline
 & & & & & \vspace{-0.40cm} \\
$j^\mathcal{P}$ & $\mu_\mathrm{q} = 0.0040$ & $\mu_\mathrm{q} = 0.0064$ & $\mu_\mathrm{q} = 0.0085$ & $\mu_\mathrm{q} = 0.0100$ & $\mu_\mathrm{q} = 0.0150$ \\
 & & & & & \vspace{-0.40cm} \\
\hline
 & & & & & \vspace{-0.41cm} \\
\hline
 & & & & & \vspace{-0.40cm} \\
$(1/2)^{-,\ast} \ \equiv \ S^\ast$ &  $777(17)$ &  $808(19)$ &  $839(22)$ &  $780(34)$ &  $782(32)$ \\
$(1/2)^+ \ \equiv \ P_-$           &  $389(16)$ &  $428(12)$ &  $447(10)$ &  $456(17)$ &  $495(16)$ \\
 & & & & & \vspace{-0.40cm} \\
\hline
 & & & & & \vspace{-0.40cm} \\
$(3/2)^+ \ \equiv \ P_+$           &  $473(10)$ &   $496(8)$ &   $488(7)$ &  $486(12)$ &  $479(14)$ \\
$(3/2)^- \ \equiv \ D_\pm$           &  $813(24)$ &  $828(19)$ &  $833(16)$ &  $861(27)$ &  $858(21)$ \\
 & & & & & \vspace{-0.40cm} \\
\hline
 & & & & & \vspace{-0.40cm} \\
$(5/2)^- \ \equiv \ D_+$           &  $823(24)$ &  $887(14)$ &  $887(15)$ &  $862(24)$ &  $846(42)$ \\
$(5/2)^+ \ \equiv \ F_\pm$           & $1134(35)$ & $1205(27)$ & $1173(24)$ & $1136(34)$ & $1205(28)$\vspace{-0.40cm} \\
 & & & & & \\
\hline
\end{tabular}
\caption{\label{TAB005}static-light mass differences $m(j^\mathcal{P}) - m(S)$ in MeV for different $\mu_\mathrm{q}$.}
\end{center}
\end{table}

To check the stability of the fitting method, we have performed
computations with different parameters (number of states $n$, fitting
range $t_\textrm{min} \ldots t_\textrm{max}$, operator content of the
correlation matrices). We have obtained results which are consistent
within statistical errors.


\subsection{Extrapolation to physical light quark masses}

We linearly extrapolate our static-light mass differences in
$(m_\mathrm{PS})^2$ to the physical $u/d$  quark mass ($m_\mathrm{PS} =
135 \, \textrm{MeV}$) and the physical $s$ quark mass (taken here as
$m_\mathrm{PS} = 700 \, \textrm{MeV}$).  Results are shown in
Figure~\ref{FIG004} and Table~\ref{TAB006}. We also list the
corresponding $\chi^2 / \mathrm{dof}$  values indicating that straight
lines are acceptable for extrapolation. A more thorough study using 
extrapolations based on chiral effective theories will be attempted 
when we are able to extract the continuum limit of our results at each 
light quark mass value. 

\begin{table}[h!]
\begin{center}
\begin{tabular}{|c|c|c|c|}
\hline
 & & & \vspace{-0.40cm} \\
 & $u/d$ quark extrapolation: & $s$ quark extrapolation: & \\
$j^\mathcal{P}$ & $m(j^\mathcal{P}) - m(S)$ in MeV & $m(j^\mathcal{P}) - m(S)$ in MeV & $\chi^2 / \mathrm{dof}$ \\
 & & & \vspace{-0.40cm} \\
\hline
 & & & \vspace{-0.41cm} \\
\hline
 & & & \vspace{-0.40cm} \\
$(1/2)^{-,\ast} \ \equiv \ S^\ast$ & $791(23)$ & $816(43)$ & $1.82$ \\
$(1/2)^+ \ \equiv \ P_-$ & $371(16)$ & $554(23)$ & $0.44$ \\
 & & & \vspace{-0.40cm} \\
\hline
 & & & \vspace{-0.40cm} \\
$(3/2)^+ \ \equiv \ P_+$ & $487(11)$ & $486(19)$ & $1.22$ \\
$(3/2)^- \ \equiv \ D_\pm$ & $804(23)$ & $887(33)$ & $0.21$ \\
 & & & \vspace{-0.40cm} \\
\hline
 & & & \vspace{-0.40cm} \\
$(5/2)^- \ \equiv \ D_+$ & $864(27)$ & $894(50)$ & $2.24$ \\
$(5/2)^+ \ \equiv \ F_\pm$ & $1149(33)$ & $1215(44)$ & $1.40$\vspace{-0.40cm} \\
 & & & \\
\hline
\end{tabular}
\caption{\label{TAB006}static-light mass differences linearly extrapolated to the physical $u/d$ quark mass and the physical $s$ quark mass.}
\end{center}
\end{table}

Note that we consider the unitary sector, where valence quarks and sea
quarks are of the same mass.  For the $s$ quark extrapolated results
this implies a sea of two degenerate $s$ instead of a sea of $u$ and
$d$. If the sea-quark mass dependence of our spectra is small, as  
usually assumed, then our results will be a good estimate of the
physical  static-strange meson spectrum. This limitation can be
removed, in principle,  by performing similar computations on $N_f =
2+1+1$ flavour gauge configurations,  which are currently being produced
by ETMC \cite{Chiarappa:2006ae}.

We have performed a similar extrapolation for the mass difference of the
$P$ wave states. When extrapolating to the physical $u/d$ quark mass, we
find $m(P_+) - m(P_-) =  117(17) \, \textrm{MeV}$, i.e.\ the $P_- \equiv
(1/2)^+$ state is lighter than  the $P_+ \equiv (3/2)^+$ as usually
expected. When increasing the mass of the light quark, we observe a
reversal of this level ordering, $m(P_-) - m(P_+) = 71(23) \,
\textrm{MeV}$ at the physical $s$ quark mass. It will be interesting to
study this in the continuum limit, in particular since such a reversal
is predicted by certain phenomenological models
\cite{Schnitzer:1978gq,Schnitzer:1989xr,Ebert:1997nk,Isgur:1998kr}.

In principle, our excited states could be two-particle states since we
have dynamical sea quarks. In practice, the two-particle state is
expected to be weakly coupled to the operators we use (which are
constructed assuming one particle states). Some exploration of
transitions to two particle static-light mesons has been made which
confirms this expectation \cite{McNeile:2004rf}.


\section{\label{SEC576}Predictions for $B$ and $B_s$ mesons}

To make predictions regarding the spectrum of $B$ and $B_s$ mesons, we
interpolate between the static-light lattice results obtained in the
previous section and experimental results for charmed mesons\footnote{For the states $B$, $D$, $D^\ast$, $D_0^\ast$ and $D_2^\ast$ experimental results for charged as well as for uncharged mesons exist. We use the average in the following.} \cite{PDG}.
To this end, we assume a linear dependence in $1 / m_Q$, where $m_Q$ is
the mass of the heavy quark. This interpolation introduces a possible
systematic error, which, however, we consider to be smaller than  the
systematic errors coming from the continuum limit, the extrapolation  to
light quarks and the treatment of the strange sea. The most important 
of these systematic errors is that involved in the continuum limit and
that will be  quantified when we have results at finer lattice spacings.


\subsection{$B$ mesons}

Results of the interpolation between our $u/d$ extrapolated $P$ wave
lattice results and experimental results on $D$ mesons are shown in
Figure~\ref{FIG005}a and Table~\ref{TAB007}.
\begin{itemize}
\item To predict $m(B_0^\ast) - m(B)$ and $m(B_1^\ast) - m(B)$, we interpolate between our static spin degenerate $P_- \equiv (1/2)^+$ state, i.e.\ $m(P_-) - m(S)$, and experimental data on \\ $m(D_0^\ast) - m(D)$ and $m(D_1(2430)^0) - m(D)$.

\item To predict $m(B_1) - m(B)$ and $m(B_2^\ast) - m(B)$, we interpolate between our static spin degenerate $P_+ \equiv (3/2)^+$ state, i.e.\ $m(P_+) - m(S)$, and experimental data on \\ $m(D_1(2420)^0) - m(D)$ and $m(D_2^\ast) - m(D)$. Here we assign the $D_1^0$ states assuming that states with similar widths belong to the same multiplet.

\item The line labeled ``$S \equiv (1/2)^-$'' in Figure~\ref{FIG005}a shows that $m(B^\ast) - m(B)$ is lighter by a factor of $\approx m_c / m_b$ than $m(D^\ast) - m(D)$ indicating that a straight line is a suitable ansatz for interpolation and that the estimate of $m_c/ m_b=0.3$ \cite{PDG} is reasonable.

\item A comparison with experimental results from CDF and D{\O} \cite{Mommsen:2006ai,Abazov:2007vq} on $m(B_1) - m(B)$ and $m(B_2^\ast) - m(B)$ shows that our lattice results are larger by $\approx 10 \%$ (cf.\ Table~\ref{TAB007}). There is another resonance listed in \cite{PDG} with unknown quantum numbers $J^\mathcal{P}$, $m(B_J^\ast) - m(B)$, which is rather close to our $m(B_0^\ast) - m(B)$ and $m(B_1^\ast) - m(B)$ results. For a conclusive comparison it will be necessary to study the continuum limit, which will be part of an upcoming publication.
\end{itemize}

\begin{figure}[h!]
\begin{center}
\input{FIG005_.pstex_t}
 \caption{\label{FIG005}Static-light mass differences linearly
interpolated to the physical $b$ quark mass.}
 \end{center}
\end{figure}

\begin{table}[h!]
\begin{center}
\begin{tabular}{|c|c|c|c|c||c|c|c|c|c|}
\hline
 & \multicolumn{4}{c||}{\vspace{-0.40cm}} & & \multicolumn{4}{c|}{} \\
 & \multicolumn{4}{c||}{$m-m(B)$ in MeV} & & \multicolumn{4}{c|}{$m-m(B_s)$ in MeV} \\
 & \multicolumn{4}{c||}{\vspace{-0.40cm}} & & \multicolumn{4}{c|}{} \\
\hline
 & & & & & & & & & \vspace{-0.40cm} \\
 state & lattice & CDF & D{\O} & PDG & state & lattice & CDF & D{\O} & PDG \\
 & & & & & & & & & \vspace{-0.40cm} \\
\hline
 & & & & & & & & & \vspace{-0.40cm} \\
 $B_0^\ast$ & $413(19)$ & & & & $B_{s0}^\ast$ & $493(16)$ & & & \\
 $B_1^\ast$ & $428(19)$ & & & & $B_{s1}^\ast$ & $535(16)$ & & & \\
 $B_1$ & $508(8)$ & $454(5)$ & $441(4)$ & & $B_{s1}$ & $510(13)$ & $463(1)$ & & \\
 $B_2^\ast$ & $519(8)$ & $458(6)$ & $467(4)$ & & $B_{s2}^\ast$ & $521(13)$ & $473(1)$ & $473(2)$ & \\
 & & & & & & & & & \vspace{-0.40cm} \\
\hline
 & & & & & & & & & \vspace{-0.40cm} \\
 $B_J^\ast$ & & & & $418(8)$ & $B_{sJ}^\ast$ & & & & $487(16)$\vspace{-0.40cm} \\
 & & & & & & & & & \\
\hline
\end{tabular}
 \caption{\label{TAB007}lattice and experimental results for $P$ wave
$B$ and $B_s$ states. Errors on lattice results are statistical only.}
 \end{center}
\end{table}


\subsection{$B_s$ mesons}

For $B_s$ mesons we proceed in the same way as for $B$ mesons, using our
$s$ quark extrapolated static-light lattice results and experimental
results on $D_s$ mesons (cf.\ Figure~\ref{FIG005}b and
Table~\ref{TAB007}).
\begin{itemize}
 \item To predict $m(B_{s 0}^\ast) - m(B_s)$ and $m(B_{s 1}^\ast) -
m(B_s)$, we interpolate between our static  spin degenerate $P_- \equiv
(1/2)^+$ state, i.e.\ $m(P_-) - m(S)$, and experimental data on \\
$m(D_{s 0}^\ast) - m(D_s)$ and $m(D_{s 1}(2460)) - m(D_s)$.

\item To predict $m(B_{s 1}) - m(B_s)$ and $m(B_{s 2}^\ast) - m(B_s)$, we interpolate between our static spin degenerate $P_+ \equiv (3/2)^+$ state, i.e.\ $m(P_+) - m(S)$, and experimental data on \\ $m(D_{s1}(2536)) - m(D_s)$ and $m(D_{s 2}) - m(D_s)$. This time we assign the $D_{s1}$ states according to the expectation that the splitting between $D_{s 1}(\textrm{``}j=3/2\textrm{''})$ and $D_{s 2}$ is roughly \\ $m_b / m_c \approx 3.3$ times larger than that between $B_{s 1}$ and $B_{s 2}^\ast$, which is according to \cite{:2007tr} approximately $10 \, \textrm{MeV}$. We also illustrate the opposite assignment in Figure~\ref{FIG005}b for completeness.

\item The line labeled ``$S \equiv (1/2)^-$'' in Figure~\ref{FIG005}b shows that $m(B_s^\ast) - m(B_s)$  is lighter by a factor of $\approx m_c / m_b$ than $m(D_s^\ast) - m(D_s)$ indicating that a straight line is a suitable ansatz for interpolation and that the estimate of $m_c/m_b=0.3$ \cite{PDG} is reasonable.

\item A comparison with experimental results from CDF and D{\O} \cite{:2007tr,:2007sna} on $m(B_1) - m(B^0)$ and $m(B_2^\ast) - m(B^0)$ shows that our lattice results are larger by $\approx 10 \%$ (cf.\ Table~\ref{TAB007}). There is another resonance listed in \cite{PDG} with unknown quantum numbers $J^\mathcal{P}$, $m(B_{s J}^\ast) - m(B_s)$, which is rather close to our $m(B_{s 0}^\ast) - m(B_s)$ result. For a conclusive comparison it will be necessary to study the continuum limit, which will be part of an  upcoming publication.

\item We also plot the $B K$ and $B^\ast K$ thresholds in
Figure~\ref{FIG005}b.  The fact that our lattice results on the $P$ wave
states $B_{s 0}^\ast$, $B_{s 1}^\ast$, $B_{s 1}$ and $B_{s 2}^\ast$  are
larger indicates that corresponding decays are energetically allowed. 
Therefore, one should expect that these states may have a larger width
compared to the corresponding excited $D_s$ states. 
\end{itemize}


\section{\label{SEC488}Conclusions}

We have explored the low lying static-light meson spectrum using $N_f =
2$ flavours of sea quarks with Wtm lattice QCD. We have presented
results for total angular momentum of the light degrees of freedom $j = 1/2$, $j = 3/2$ and
$j = 5/2$ and for parity $\mathcal{P} = +$ and $\mathcal{P} = -$. The
lattice spacing is $a = 0.0855(5) \, \textrm{fm}$ and we have
considered five different values for the light quark mass corresponding
to $300 \, \textrm{MeV} \ltapprox m_\mathrm{PS} \ltapprox 600 \,
\textrm{MeV}$.

We have extrapolated our results in $(m_\mathrm{PS})^2$ both to the
physical $u/d$  quark mass and to the physical $s$ quark mass. Moreover,
we used experimental results from  $D$ and $D_s$ mesons to interpolate
in the heavy quark mass from the static case to the physical $b$ quark
mass.  We are able to predict the spectrum of excited $B$ and $B_s$ 
mesons from first principles. Our formalism has lattice artifacts of 
order $a^2$ and we shall be able to control these in future work by 
studying smaller $a$ values. Comparing our current predictions to available experimental results, we find agreement up to $10\%$ with $P$ wave $B$ and $B_s$ mesons.

Throughout this paper we have considered the unitary sector, where
valence quarks and sea quarks are of the same mass. Particularly for our
$B_s$ results, this implies a sea of two degenerate $s$ instead of a sea
of $u$ and $d$. We plan to improve this by performing similar
computations on $N_f = 2+1+1$ flavour gauge configurations, which are
currently produced by ETMC. Another important issue in the near future
will be an investigation of the continuum limit, which amounts to
considering other values for the lattice spacing. Such a study will be
necessary  for a conclusive comparison between lattice results and
experimental results for $B$ and $B_s$ mesons. We also plan to compute
static-light decay constants and to make a detailed comparison with
chiral effective Lagrangians.


\section*{Acknowledgments}

MW would like to thank Carsten Urbach for help in retrieving and handling ETMC gauge configurations. Moreover, we acknowledge useful discussions with Benoit Blossier, Tommy Burch, Vladimir Galkin, Christian Hagen, Rainer Sommer and Carsten Urbach. This work has been supported in part by the DFG Sonderforschungsbereich/Transregio SFB/TR9-03.



\end{document}